\documentclass[prc,aps,twocolumn,superscriptaddress,nofootinbib,showpacs,showkeys]{revtex4}

\usepackage{graphicx,epsfig,latexsym,amssymb,amsmath,amsthm,color,dcolumn,bm}
\usepackage[bookmarksnumbered,plainpages]{hyperref}
\usepackage{booktabs,longtable}

\begin{document}

\title{A new barrier penetration formula and its application to $\alpha$-decay half-lives}

\author{Lu-Lu Li}
 \affiliation{Key Laboratory of Frontiers in Theoretical Physics,
              Institute of Theoretical Physics, Chinese Academy of Sciences, Beijing 100190, China}
\author{Shan-Gui Zhou}
 \email{sgzhou@itp.ac.cn}
 \homepage{http://www.itp.ac.cn/~sgzhou}
 \affiliation{Key Laboratory of Frontiers in Theoretical Physics,
              Institute of Theoretical Physics, Chinese Academy of Sciences, Beijing 100190, China}
 \affiliation{Center of Theoretical Nuclear Physics, National Laboratory
              of Heavy Ion Accelerator, Lanzhou 730000, China}
\author{En-Guang Zhao}
 \affiliation{Key Laboratory of Frontiers in Theoretical Physics,
              Institute of Theoretical Physics, Chinese Academy of Sciences, Beijing 100190, China}
 \affiliation{School of Physics, Peking University,
              Beijing 100871, China}
 \affiliation{Center of Theoretical Nuclear Physics, National Laboratory
              of Heavy Ion Accelerator, Lanzhou 730000, China}
\author{Werner Scheid}
 \affiliation{Institute for Theoretical Physics of Justus-Liebig-University,
              35392 Giessen, Germany}

\date{\today}

\begin{abstract}
Starting from the WKB approximation, a new barrier penetration
formula is proposed for potential barriers containing a long-range
Coulomb interaction. This formula is especially proper for the
barrier penetration with penetration energy much lower than the
Coulomb barrier. The penetrabilities calculated from the new formula
agree well with the results from the WKB method. As a first attempt,
this new formula is used to evaluate $\alpha$ decay half-lives of
atomic nuclei and a good agreement with the experiment is obtained.
\end{abstract}



\date{\today}

\maketitle

\section{\label{sec:intro}Introduction}

As a common quantum phenomenon, the tunneling through a potential
barrier plays a very important role in the microscopic world and has
been studied extensively since the birth of quantum mechanics. One
of the earliest applications of quantum tunneling is the explanation
of $\alpha$ decays in atomic nuclei. The quantum tunneling effect
governs also many other nuclear processes such as fission and
fusion. In particular, a lot of new features are revealed in
sub-barrier fusion reactions which are closely connected with the
tunneling phenomena~\cite{Dasgupta1998, Liang2003, Liang2006,
Umar2007}.

For most of the potential barriers, the penetrability can not be
calculated analytically~\cite{Aleixo2000}. Among those potentials
for which analytical solutions can be obtained, the parabolic
potential~\cite{Kemble1935, Hill1953} is the mostly used in the
study of nuclear fusion. By approximating the Coulomb barrier to a
parabola, Wong derived an analytic expression for the fusion cross
section~\cite{Wong1973} which is widely adopted today in the study
of heavy ion reactions (see, e.g., recent Refs.~\cite{Wang2008,
Smolanczuk2008}). The parabolic approximation works remarkably well
both for the penetrability and for the fusion cross section at
energies around or above the Coulomb barrier~\cite{Hagino1998}.

Apparently the parabolic approximation breaks down at energies much
smaller than the barrier height due to the long-range Coulomb
interaction. One may calculate the penetration probability
numerically by using the path integral method or the WKB
approximation. However, it is highly desirable to have an analytical
expression for the barrier penetrability when one introduces an
energy-dependent one-dimensional potential
barrier~\cite{Mohanty1990} or barrier distribution
functions~\cite{Rowley1991, Zagrebaev2001, Feng2006, Feng2007,
Wang2008c}.

In the present work, we derived a new barrier penetration formula
based on the WKB approximation. The influence of the long Coulomb
tail in the barrier potential is taken into accout properly.
Therefore this formula is especially applicable to the barrier
penetration with penetration energy much lower than the Coulomb
barrier.

As a first attempt and a test study, we apply this new formula to
evaluate $\alpha$ decay half-lives of atomic nuclei. For the
$\alpha$ decay, the penetrability is usually calculated with the WKB
approach~\cite{Buck1996, Xu2005, Denisov2005}, in other words,
integrating numerically the wave number within two turning points at
which the interaction potential is equal to the $Q$-value of the
$\alpha$ decay. We will show that the present analytical formula
reproduces the experimental results very well, especially for
spherical nuclei.

The paper is organized as follows. In Sec.~\ref{sec:formalism} we
present the new barrier penetration formula. The validity of the new
formula is investigated and its application to $\alpha$ decays are
given in Sec.~\ref{sec:results}. Finally in Sec.~\ref{sec:summary}
we summarize our work. In the Appendix, the detailed derivation of
the new penetration formula is given.

\section{\label{sec:formalism}Formalism}

When the penetration energy is well below the Coulomb barrier, the
barrier penetrability formula derived from the WKB approximation
reads,
\begin{equation}
 P(E) =
  \exp \left[ -2 \int_{ R_\mathrm{in} }^{ R_\mathrm{out} }
              \sqrt{ \frac{2\mu }{ \hbar^2 } \left( V(R) - E \right) }\ dR
       \right] ,
 \label{eq:WKB}
\end{equation}
where the potential usually consists of three parts, the nuclear,
the Coulomb, and the centrifugal potentials,
\begin{equation}
 V(R) = V_\mathrm{N}(R) + V_\mathrm{C}(R) + \frac{L(L+1)}{2\mu R^2} .
 \label{eq:potential}
\end{equation}
$R_\mathrm{in}$ and $R_\mathrm{out}$ are the inner and outer turning
points determined by the relation $V(R)=E$.

By approximating $V(R)$ to a parabola with the height $V_\mathrm{B}$
and the width $\hbar\omega$, Eq.~(\ref{eq:WKB}) is reduced as
\begin{equation}
 P(E) = \exp \left[ - \frac{2 \pi}{\hbar\omega} ( V_\mathrm{B} - E ) \right] ,
 \label{eq:HW}
\end{equation}
which has been widely used in the study of heavy ion reactions.

Because of the long-range Coulomb interaction, the Coulomb barrier
given in Eq.~(\ref{eq:potential}) has a long tail and is asymmetric.
Thus for the penetration well below the barrier, the parabolic
approximation is not valid. We may divide the potential barrier into
two parts at the barrier position $R_\mathrm{B}$. The first part of
$V(R)$ with $R_\mathrm{in}<R<R_\mathrm{B}$ could still be
approximated by half of a parabola and we need to evaluate the
integration in Eq.~(\ref{eq:WKB}) in the range
$R_\mathrm{B}<R<R_\mathrm{out}$ only. For S wave, the integral in
Eq.~(\ref{eq:WKB}) is evaluated as,
\begin{equation}
 P(E) = \exp \left[ -( x_1 + x_2 ) \right] ,
 \label{eq:x1x2}
\end{equation}
with
\begin{eqnarray}
 x_1
 & \equiv &
 2 \int_{ R_\mathrm{in} }^{ R_\mathrm{B} }
    \sqrt{ \frac{2\mu }{ \hbar^2 } \left( V(R) - E \right) }\ dR
 \nonumber \\
 & \approx &
 \frac{ \pi } { \hbar\omega } ( V_B - E ) ,
 \label{eq:left}
\end{eqnarray}
under the parabolic approximation and
\begin{eqnarray}
 x_2
 & \equiv &
 2 \int^{R_ \mathrm{out} }_{R_\mathrm{B}}
    \sqrt{ \frac{2\mu }{ \hbar^2 } \left( V(R) - E \right) }\ dR
 \nonumber \\
 & \approx &
 2 k R_\mathrm{B}
 \left[
  \tau \left( \frac{\pi}{2} - \arcsin
  \sqrt{ \frac{1}{\tau} }
   \right) -
  \sqrt{\tau - 1}
 \right]
 \nonumber \\
 &   & \mbox{}
 +
 \frac{ k a } { \sqrt{ \tau - 1 } } \frac{V_0}{E}
 \ln [ 1 + e^{ (R_0 - R_B) / a } ]
 \label{eq:new}
 ,
\end{eqnarray}
where $k=\sqrt{2\mu E} / \hbar$ and $\tau =
V_\mathrm{C}(R_\mathrm{B}) / E$. The details of the derivation of
Eq.~(\ref{eq:new}) are given in the Appendix. It should be mentioned
that in the derivation of Eq.~(\ref{eq:new}), a Woods-Saxon form is
used for $V_\mathrm{N}(R)$.

\section{\label{sec:results}Results and discussions}

In this section, we use the new formula to study the typical barrier
penetration problem, $\alpha$ decays of atomic nuclei. The $\alpha$
decay half-life is related to the decay width $\Gamma$ by
\cite{Denisov2005, Xu2005a, Dupre2006}
\begin{equation}
 T_{1/2} = \frac{ \hbar \ln 2}{\Gamma} .
 \label{eq:halflife}
\end{equation}
The decay width $\Gamma$ is calculated as~\cite{Denisov2005}
\begin{equation}
 \Gamma = \hbar \nu S P(Q) = \hbar \xi P(Q) ,
 \label{eq:width}
\end{equation}
where $\nu$ is the assaults frequency of $\alpha$ particle on the
barrier, $S$ the spectroscopic or preformation factor and $P(Q)$ the
penetrability with $Q$ the $\alpha$ decay Q-value. For spherical
nuclei, $\xi$ is parametrized as~\cite{Denisov2005}
\begin{equation}
 \xi = ( 6.1814 + 0.2988 A^{-1/6} ) \times 10^{19} \quad \textrm{s}^{-1} ,
 \label{eq:prefomation}
\end{equation}
and the penetrability will be calculated with Eqs.~(\ref{eq:x1x2}),
(\ref{eq:left}), and (\ref{eq:new}).

\begin{figure}
\includegraphics[width=0.45\textwidth]{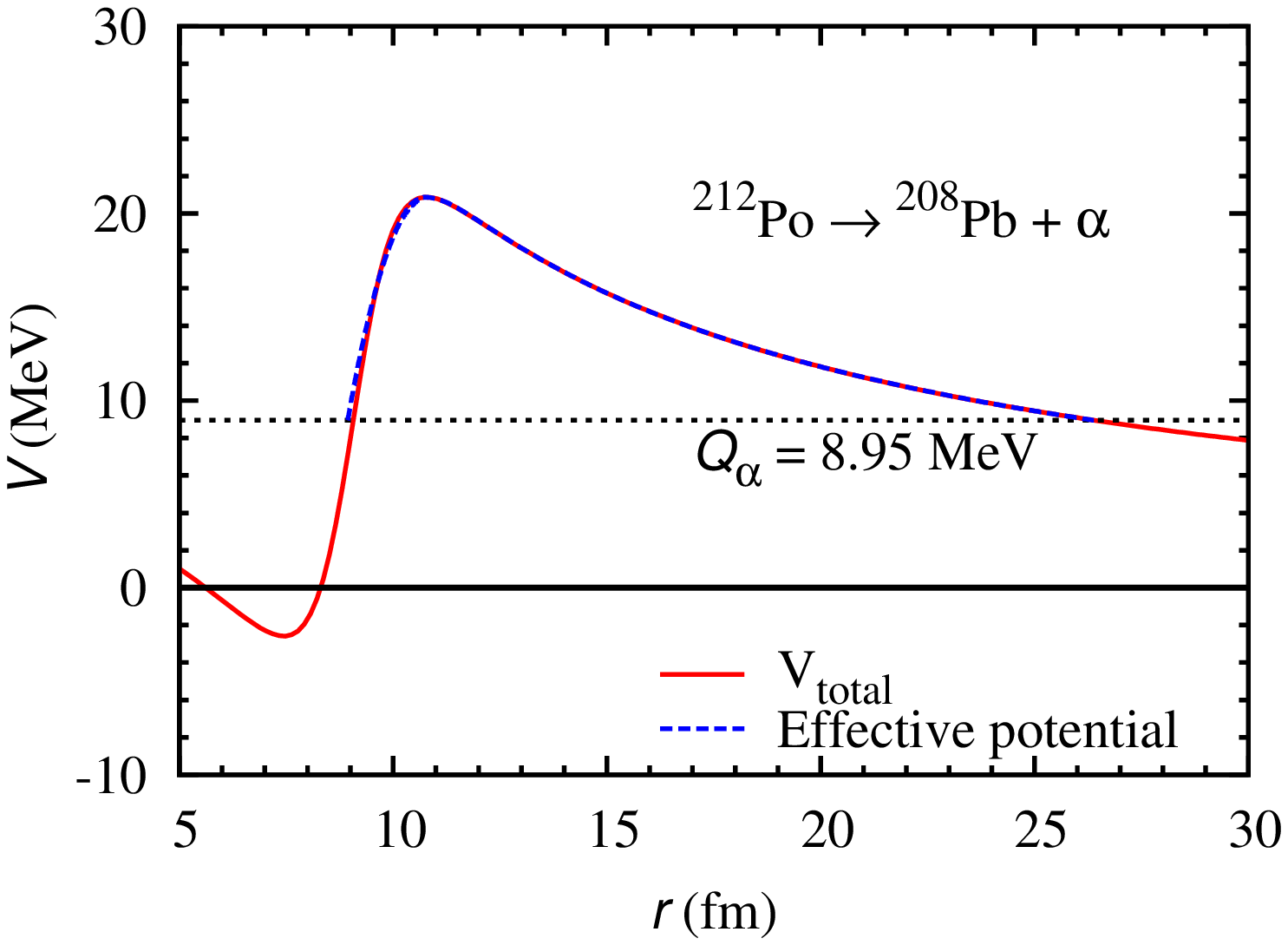}
\includegraphics[width=0.45\textwidth]{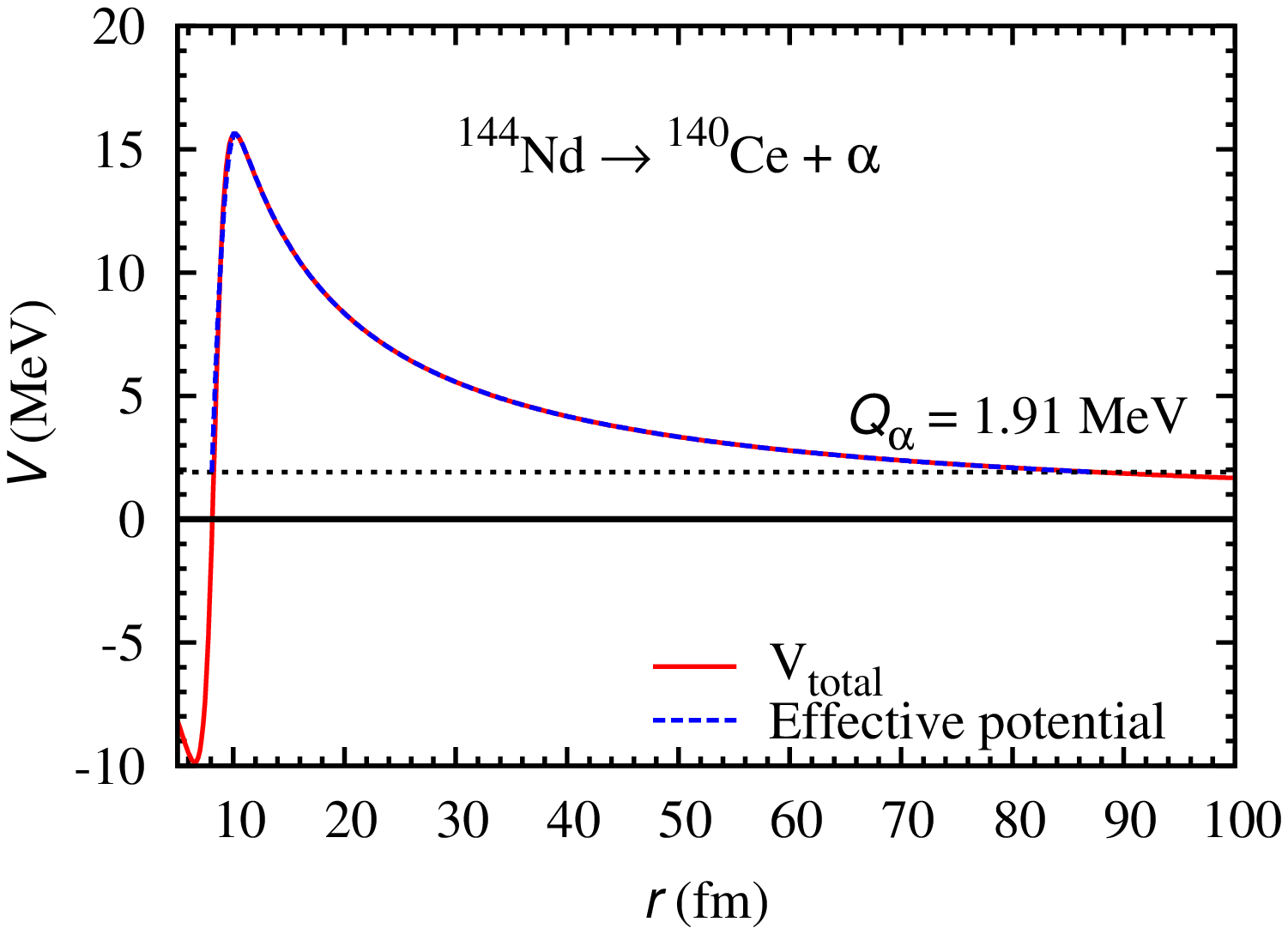}
\caption{\label{fig:potential} (Color online) The barrier potential
between the $\alpha$ and the daughter nucleus for $^{212}$Po and
$^{144}$Nd. The solid curve shows the exact potential $V(R)$ and the
dashed curve stands for the effective potential given in
Eq.~(\ref{eq:Veff}) associated with the parabolic approximation
Eq.~(\ref{eq:left}) and the new barrier penetration formula
Eq.~(\ref{eq:new}). Note that the two curves are almost identical to
each other. }
\end{figure}

For the $\alpha$-nuclear interaction, we adopt the Coulomb and the
Woods-Saxon potentials and parameters proposed in
Ref.~\cite{Denisov2005},
\begin{equation}
 V_\mathrm{C}(R) =
 \begin{cases}
  \displaystyle \frac{2Ze^2}{R} , & R \ge R_m , \\
  \displaystyle \frac{Ze^2}{R_m} \left[ 3 - \frac{R^2}{R^2_m} \right] , & R \le R_m ,
 \end{cases}
 \label{eq:Coulomb}
\end{equation}
and
\begin{eqnarray}
 V_\mathrm{N}(R)
 & = &
 \frac{V(A,Z,Q)}{1+\exp\left[(R-R_m)/a\right]}
 ,
\end{eqnarray}
with $A$ and $Z$ the mass and charge numbers of the daughter nucleus
and $Q$ the $\alpha$ decay energy. The parameters in these
potentials and given in Eq.~(\ref{eq:prefomation}) were obtained by
fitting $\alpha$ decay half lives and cross section data for several
fusion reactions. It can be easily verified that the position of the
Coulomb barrier $R_\mathrm{B}$ is larger than $R_m$ thus the use of
the Coulomb force given in Eq.~(\ref{eq:Coul}) is valid.

\subsection{Validity of the new formula}

Before the new formula is used to study alpha decays, we investigate
in details its validity. First we examine how the effective
potential connected with the new formula Eq.~(\ref{eq:new}) is close
to the exact one. Two extreme examples are chosen for this purpose,
$^{212}$Po which has a very short half-life $3.02 \times 10^{-7}$ s
and $^{144}$Nd which has a quite long half-life $7.24 \times
10^{22}$ s~\cite{Duarte2002}. The barrier potential $V(R)$ is shown
in Fig.~\ref{fig:potential} for these two systems. The effective
potential,
\begin{widetext}
\begin{equation}
 V_\mathrm{eff}(R) =
 \begin{cases}
  V_\mathrm{B} - \frac{1}{2} \mu \omega^2 (R - R_B)^2, & R_\mathrm{in}<R<R_\mathrm{B}, \\
  V_\mathrm{C}(R) +
  \displaystyle
  \frac{ V_\mathrm{C}(R) - E } { V_\mathrm{C}(R_B) - E } V_\mathrm{N}(R) +
  \frac{1}{4} \frac{ V_\mathrm{N}^2(R) }{ V_\mathrm{C}(R_\mathrm{B}) - E } ,
  & R_\mathrm{B}<R<R_\mathrm{V} , \\
  V_\mathrm{C}(R) ,
  & R_\mathrm{V}<R<R_\mathrm{out} , \\
 \end{cases}
 \label{eq:Veff}
\end{equation}
\end{widetext}
is also shown for comparison. $R_\mathrm{V} \approx 9-12$ fm is the
radial position outside of which the nuclear part of the
$\alpha$-nucleus potential could be neglected (see the Appendix for
more details). In our calculations, the width of the parabolic
potential is obtained by fitting the barrier potential from the
inner turning point $R_{ \mathrm{in} }$ to the position of the
barrier $R_\mathrm{B}$. Unlike the full parabolic approximation, the
effective potential is asymmetric and coincides with the exact
potential very well, especially the outer side of the barrier which
critically influences $\alpha$ decays.

\begin{table}
\caption{\label{tab:deviation} Comparison of the results for the
barrier penetration probability for $\alpha$ decays in Po isotopes
(charge and mass numbers of the $\alpha$ emitter are listed in the
first and the second entries). The meaning of $x_{1,2}$ is given in
Eq.~(\ref{eq:x1x2}). The superscript ``WKB'' means the penetrability
calculated from the WKB approach, ``Para'' from the parabolic
approximation in Eq.~(\ref{eq:left}), and ``New'' from the new
formulas Eq.~(\ref{eq:new}).}
\begin{ruledtabular}
\begin{tabular}{ccc ccc c}
  $Z_\mathrm{p}$ & $A_\mathrm{p}$ & $Q_{\alpha}$ (MeV) &
  $x_1^{ \mathrm{WKB} }$ &
  $x_1^{\mathrm{New}} = x_1^{\mathrm{Para}}$ &
  $x_2^{ \mathrm{WKB} }$ &  $x_2^{ \mathrm{New} }$ \\
  \hline
 84 & 190 & 7.64 & 4.9808 & 5.0816 & 34.9523 & 35.0751  \\
 84 & 191 & 7.48 & 5.0093 & 5.1146 & 36.0311 & 36.1527  \\
 84 & 192 & 7.32 & 5.0384 & 5.1482 & 37.1506 & 37.2712  \\
 84 & 193 & 7.10 & 5.0896 & 5.2031 & 38.7753 & 38.8944  \\
 84 & 194 & 7.00 & 5.0980 & 5.2165 & 39.5213 & 39.6397  \\
 84 & 195 & 6.75 & 5.1605 & 5.2823 & 41.5276 & 41.6445  \\
 84 & 196 & 6.66 & 5.1664 & 5.2931 & 42.2564 & 42.3725  \\
 84 & 197 & 6.41 & 5.2292 & 5.3592 & 44.4401 & 44.5546  \\
 84 & 198 & 6.31 & 5.2396 & 5.3743 & 45.3311 & 45.4449  \\
 84 & 199 & 6.08 & 5.2957 & 5.4338 & 47.5227 & 47.6352  \\
 84 & 200 & 5.99 & 5.3036 & 5.4463 & 48.3954 & 48.5072  \\
 84 & 201 & 5.81 & 5.3429 & 5.4894 & 50.2458 & 50.3564  \\
 84 & 202 & 5.70 & 5.3585 & 5.5093 & 51.4094 & 51.5193  \\
 84 & 203 & 5.50 & 5.4050 & 5.5594 & 53.6532 & 53.7618  \\
 84 & 204 & 5.49 & 5.3875 & 5.5468 & 53.7347 & 53.8430  \\
 84 & 205 & 5.32 & 5.4245 & 5.5875 & 55.7496 & 55.8569  \\
 84 & 206 & 5.33 & 5.4015 & 5.5693 & 55.5907 & 55.6978  \\
 84 & 207 & 5.22 & 5.4191 & 5.5908 & 56.9347 & 57.0410  \\
 84 & 208 & 5.22 & 5.4007 & 5.5769 & 56.8997 & 57.0058  \\
 84 & 210 & 5.41 & 5.3038 & 5.4892 & 54.4768 & 54.5836  \\
 84 & 212 & 8.95 & 4.1395 & 4.3264 & 26.3317 & 26.4569  \\
 84 & 213 & 8.54 & 4.2585 & 4.4498 & 28.5307 & 28.6533  \\
 84 & 214 & 7.83 & 4.4720 & 4.6690 & 32.8310 & 32.9495  \\
 84 & 215 & 7.53 & 4.5552 & 4.7559 & 34.8360 & 34.9528  \\
 84 & 216 & 6.91 & 4.7391 & 4.9445 & 39.4763 & 39.5896  \\
 84 & 218 & 6.11 & 4.9669 & 5.1798 & 46.5717 & 46.6806  \\
\end{tabular}
\end{ruledtabular}
\end{table}

In order to examine more closely the accuracy of the new formula, we
list the calculated penetration probabilities for $\alpha$ decays of
polonium isotopes in Table~\ref{tab:deviation}. The values in the
exponential of Eq.~(\ref{eq:x1x2}), $x_{1}$ and $x_2$, calculated
from the WKB approach, the parabolic approximation and the new
formula are compared. One finds good agreement between the results
from the new formula and the WKB approach. For $x_2$, the average
relative root mean square deviation is 0.28 \%. This tells that the
present formula could be used with satisfactory accuracy in the
study the barrier penetration well below the Coulomb barrier.

\subsection{$\alpha$-decay half-lives}

\begin{figure}
\includegraphics[width=0.45\textwidth]{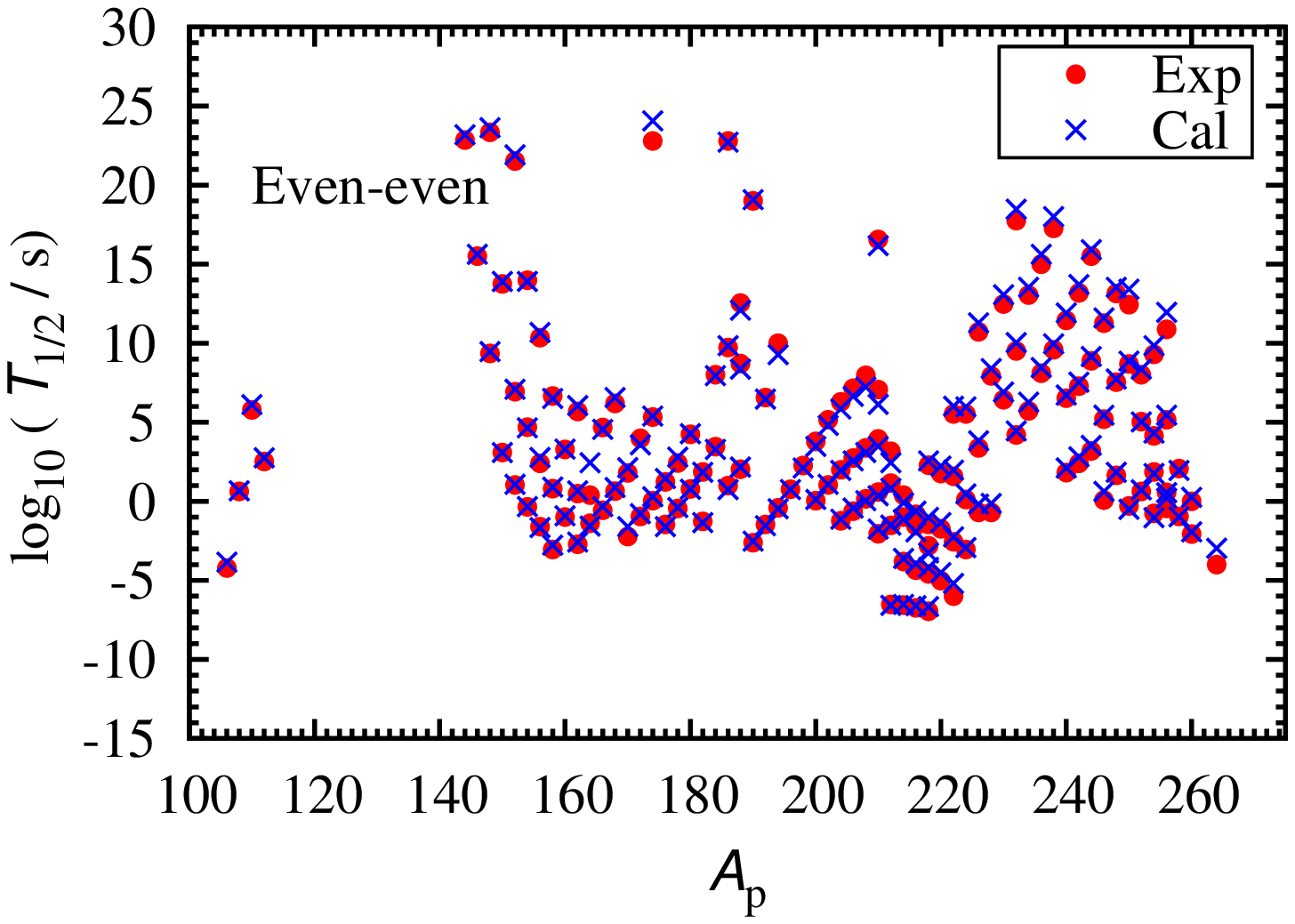}
\includegraphics[width=0.45\textwidth]{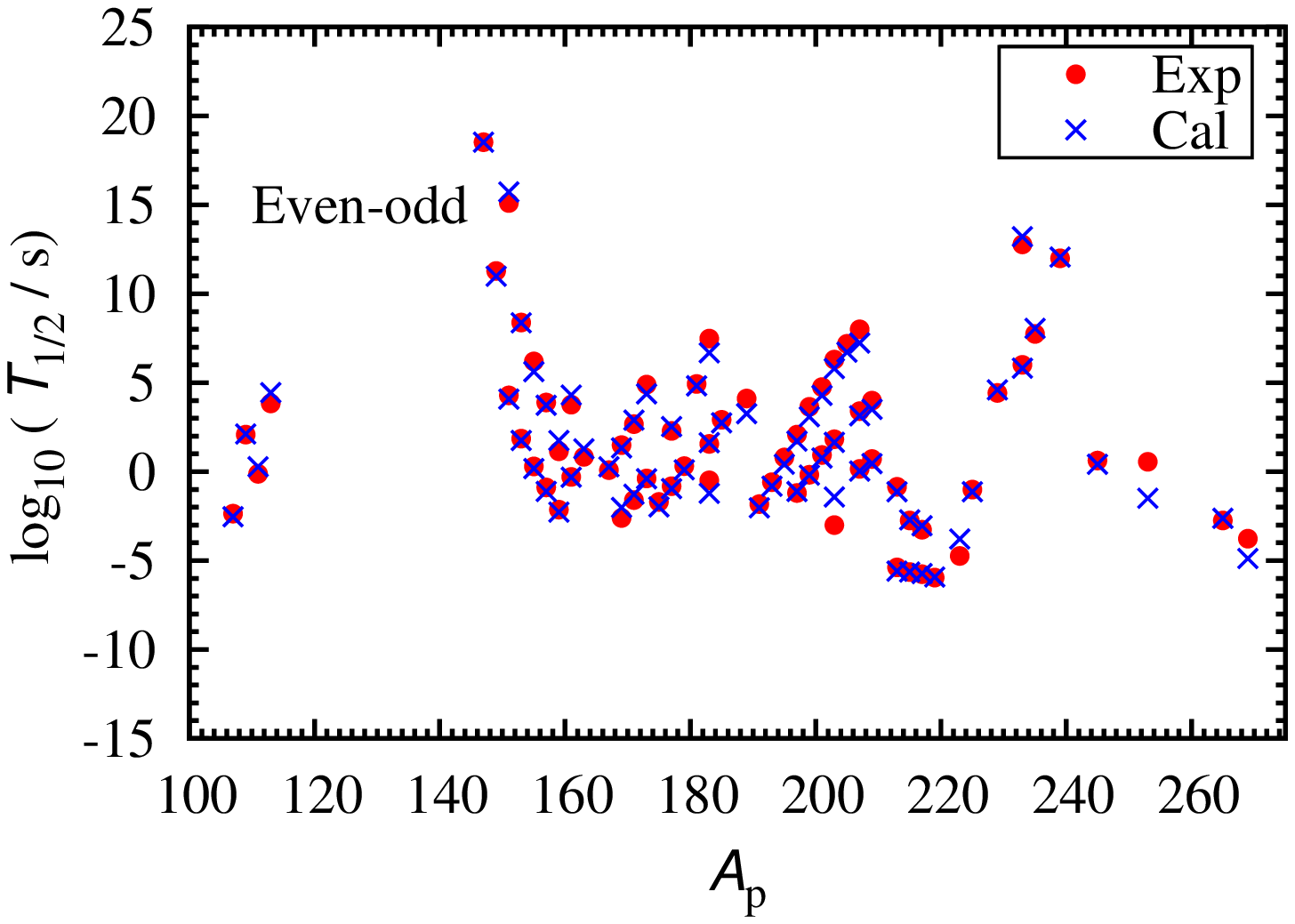}
\includegraphics[width=0.45\textwidth]{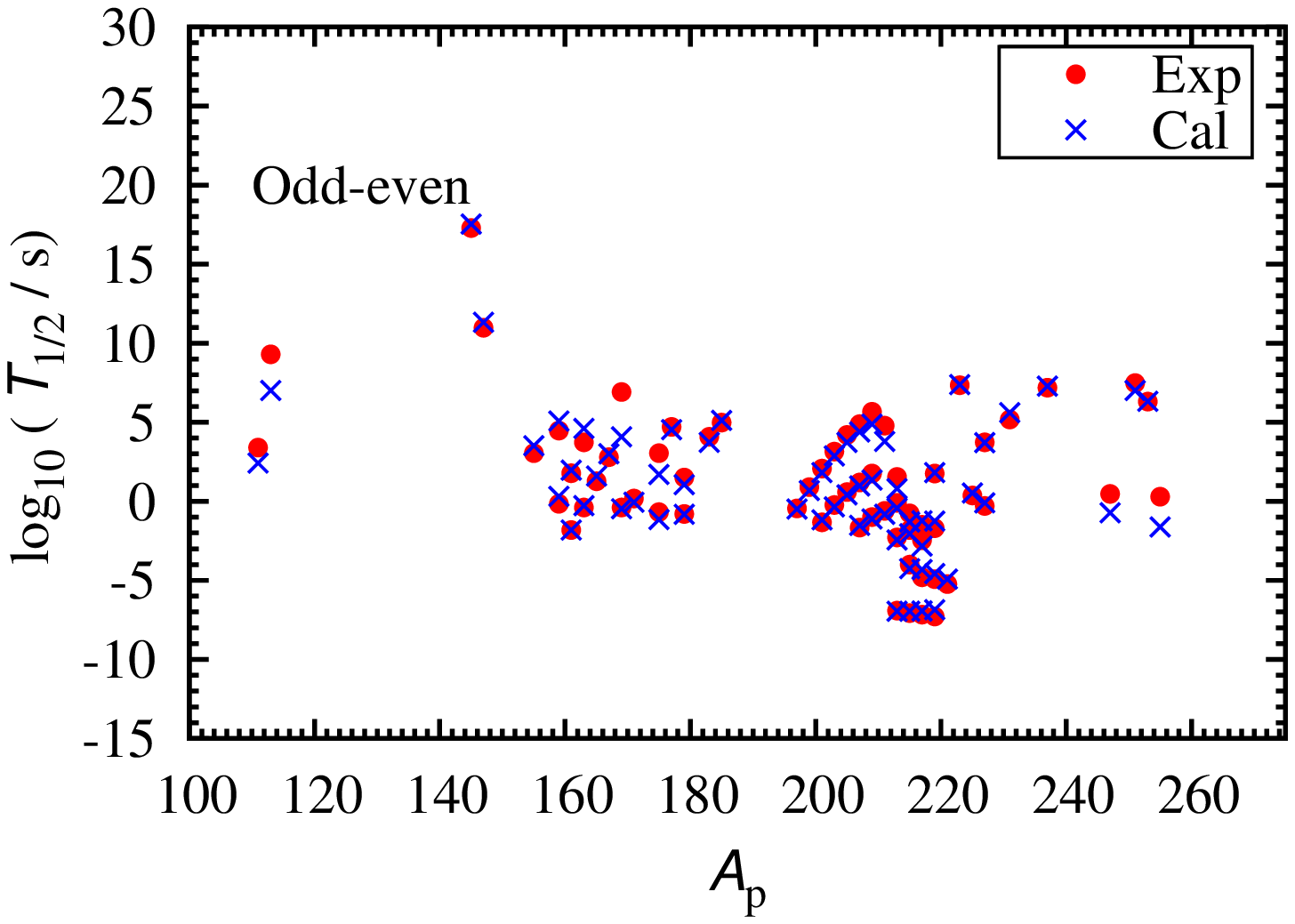}
\includegraphics[width=0.45\textwidth]{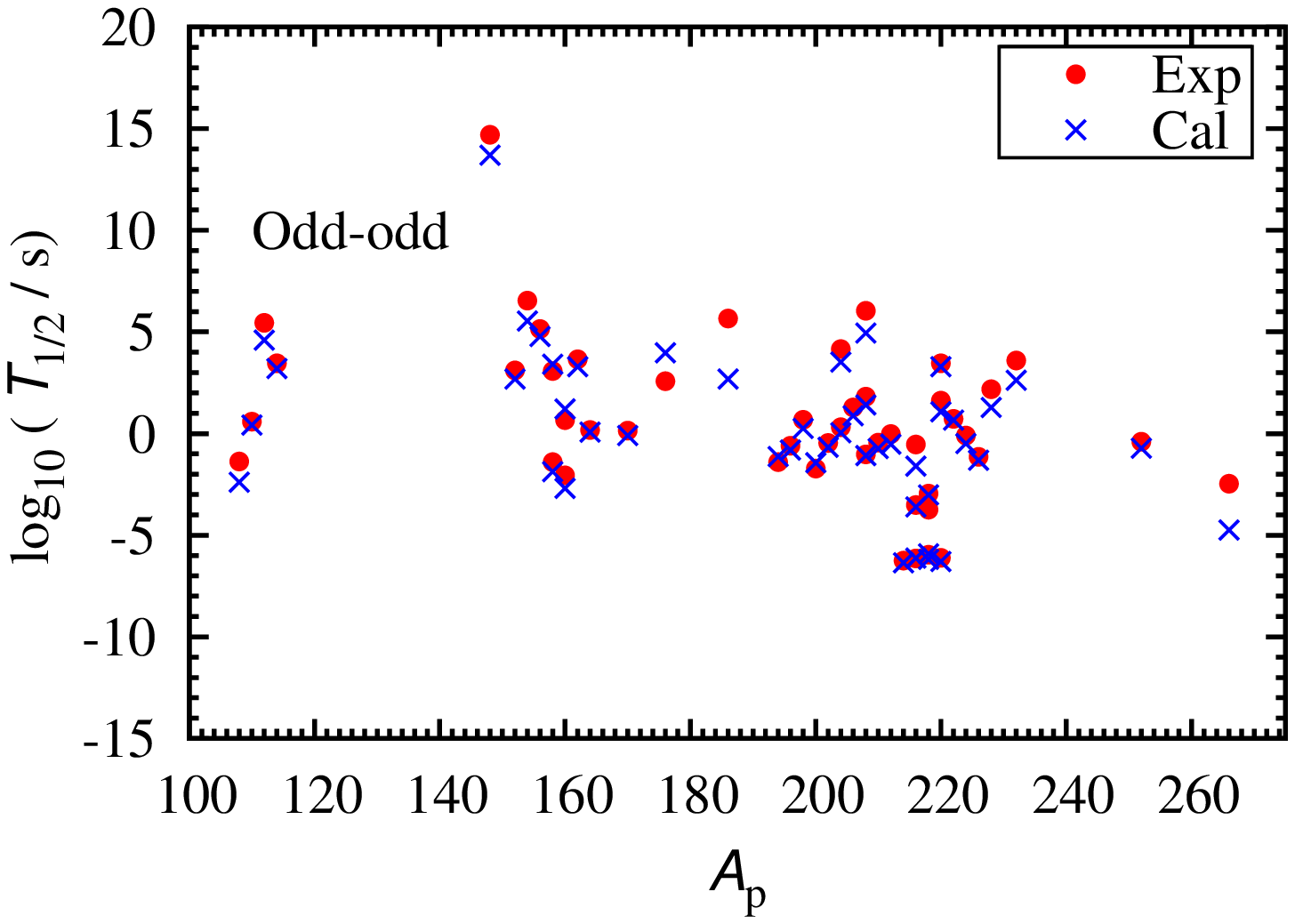}
\caption{\label{fig:compare} (Color online) Comparison of the
calculated (blue crosses) and experimental (red dots) values for
$\alpha$ decay half lives of 159 even-even, 72 even-odd, 66
odd-even, and 47 odd-odd nuclei.}
\end{figure}

\begin{figure}
\includegraphics[width=0.45\textwidth]{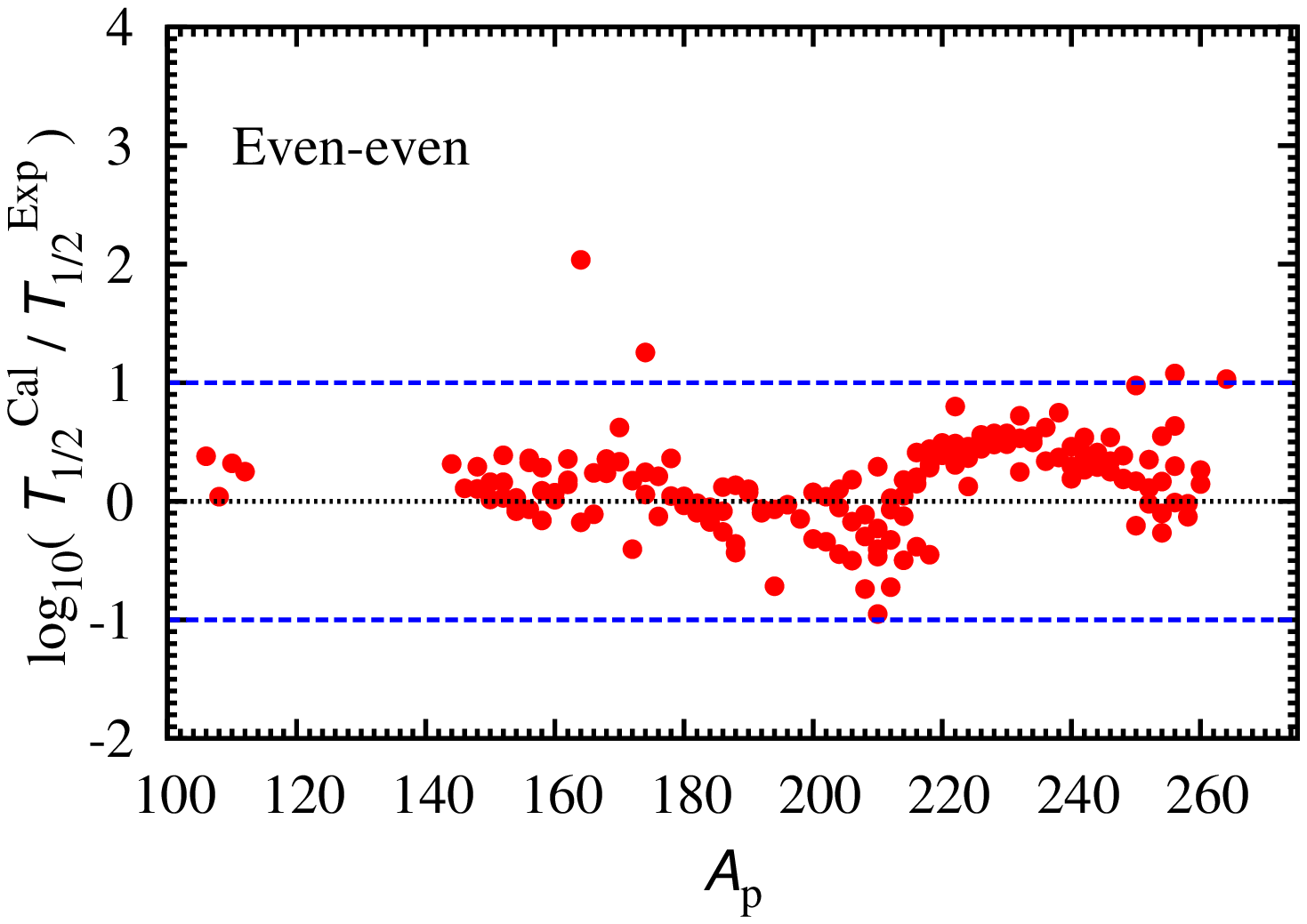}
\includegraphics[width=0.45\textwidth]{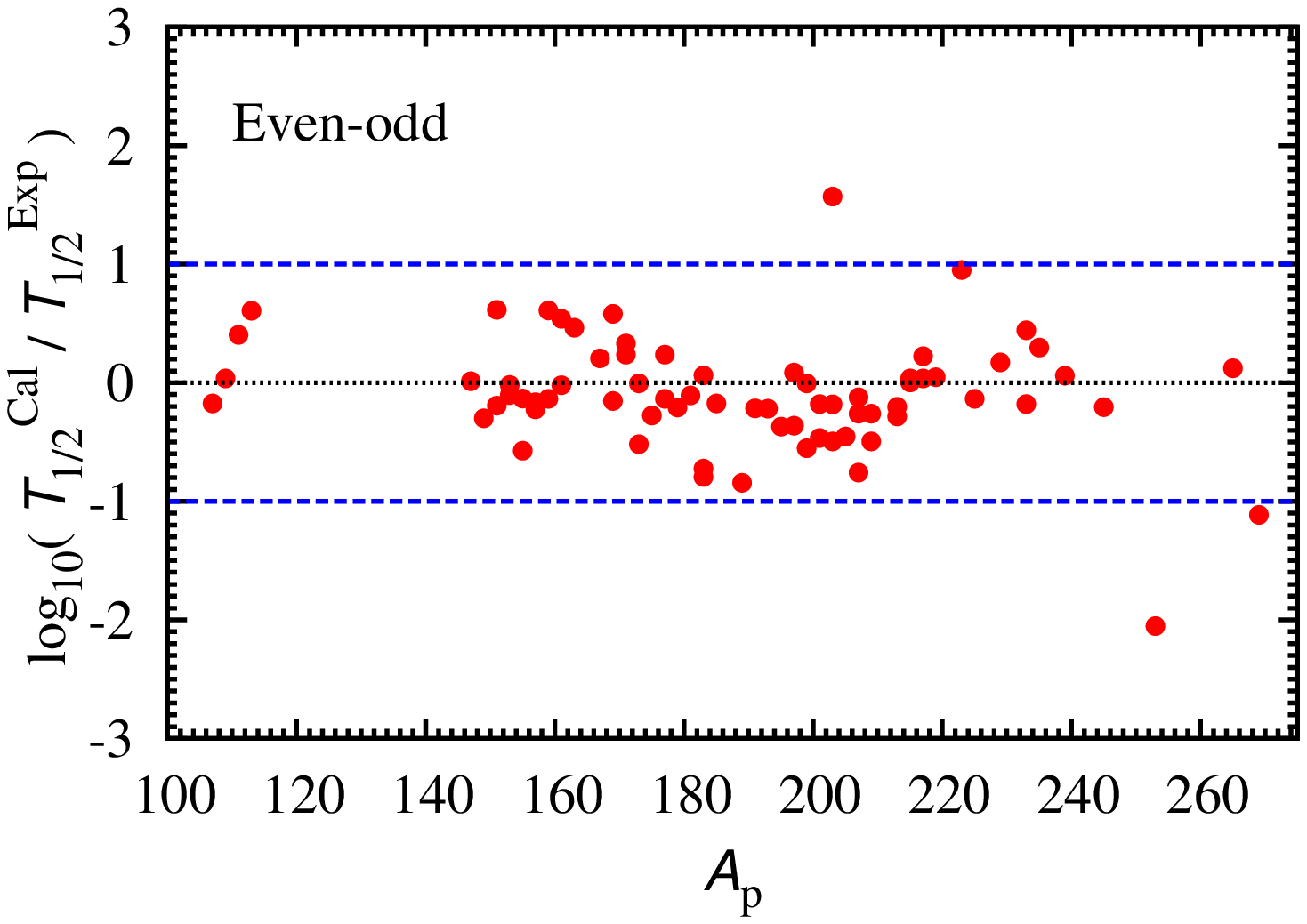}
\includegraphics[width=0.45\textwidth]{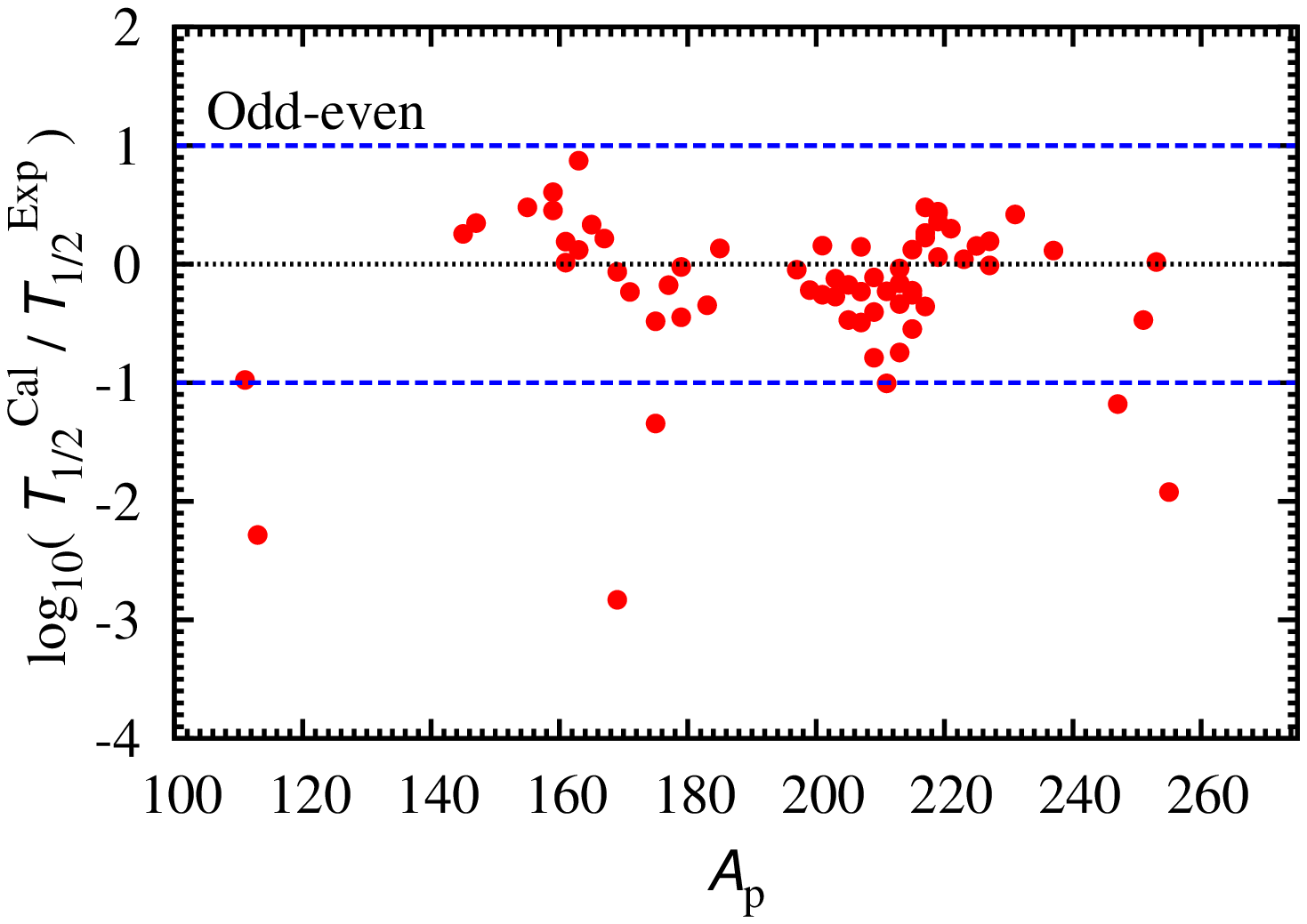}
\includegraphics[width=0.45\textwidth]{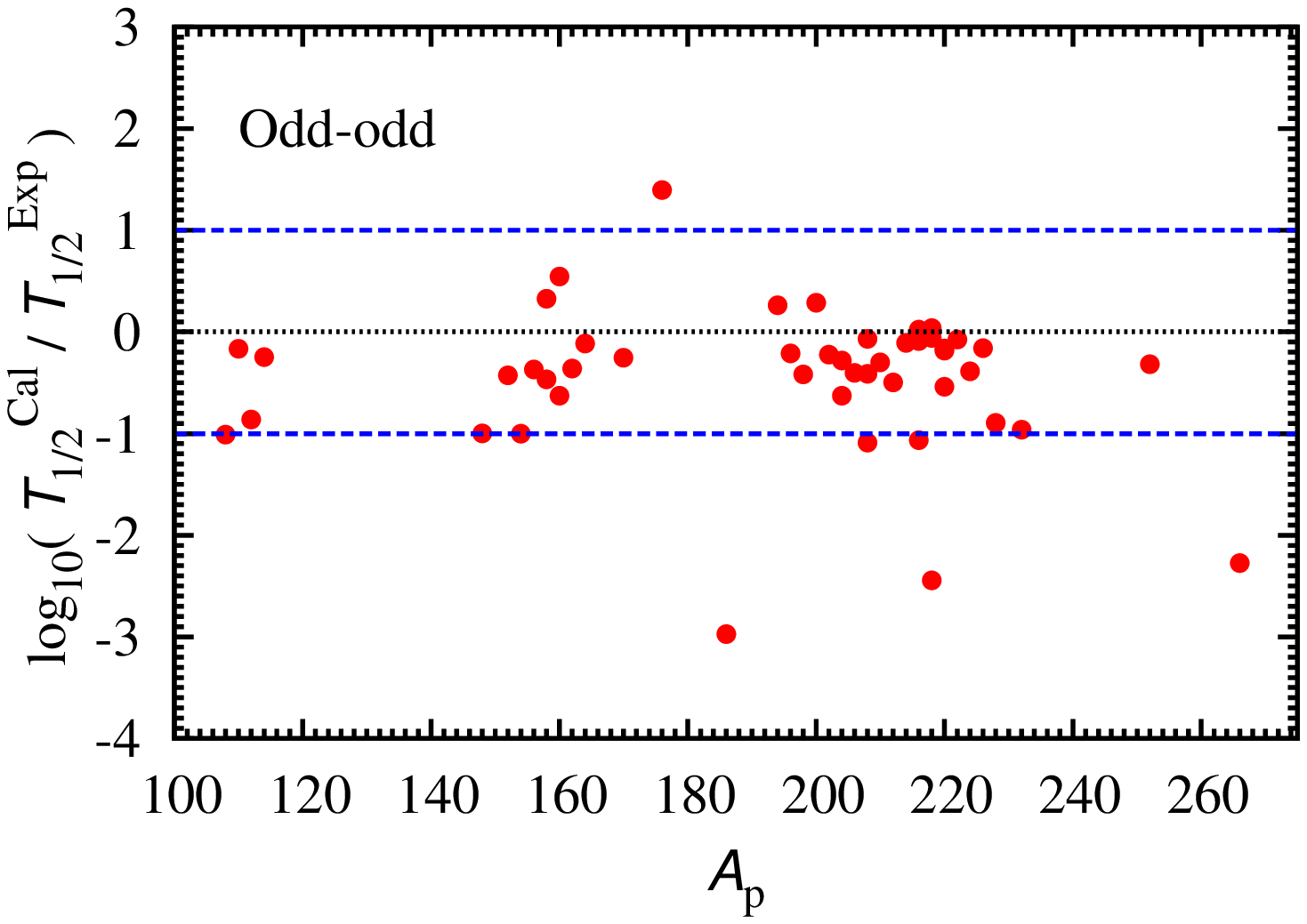}
\caption{\label{fig:ratio} (Color online) The ratios between the
calculated and experimental values for $\alpha$ decay half lives of
159 even-even, 72 even-odd, 66 odd-even, and 47 odd-odd nuclei.}
\end{figure}

The new barrier penetration formula is used to calculate
$\alpha$-decay half-lives of 344 nuclei collected in
Ref.~\cite{Duarte2002}. The experimental values of $\alpha$ decay
half lives are also taken from Ref.~\cite{Duarte2002} except for
$^{215}$Po. In Ref.~\cite{Duarte2002}, $\log(T^\mathrm{Exp}_{1/2}/s)
= -3.74$ for $^{215}$Po while in Refs.~\cite{Royer2000, Audi2003,
NWC2005} the experimental value is $\log(T^\mathrm{Exp}_{1/2}/s) =
-2.75$. We take the latter value in the present work. The
experimental values of the $\alpha$ decay half-lives range from
$10^{-7} \sim 10^{24}$ s. The $Q$ values of the $\alpha$ decays are
also taken from Ref.~\cite{Duarte2002} where these values were
calculated from the Atomic Mass Evaluation by Audi et
al.~\cite{Audi1997} or from the mass table by M\"{o}ller et
al.~\cite{Moeller1995}.

The angular momentum $L$ carried by the emitted $\alpha$ particle in
a ground-state to ground-state $\alpha$ transition of even-even
nucleus is zero. In odd-$A$ or odd-odd nuclei, $L$ could be non
zero. Because the information on $L$ is absent, in the present work
we assume $L=0$ for all $\alpha$ decays as usually
done~\cite{Denisov2005, Xu2005, Xu2005a, Zhang2006, Pei2007}.

\begin{table}
\caption{\label{tab:stati} A statistics of the ratios between the
calculated and the experimental values $S_\alpha = \log_{10}(
T^\mathrm{Cal}_{1/2} / T^\mathrm{Exp}_{1/2} )$ for the $\alpha$
decay of 344 nuclei. 68 daughter nuclei are spherical ($|\beta_2| <
0.01$) and the results for them are given in the last two lines.}
\begin{ruledtabular}
\begin{tabular}{lcc c}
 Nuclei   & $ |S_\alpha| \le 1 $ & $1 <  |S_\alpha| \le 2 $ & $ 2 < |S_\alpha| \le 3 $ \\
\hline
All       &    323     &     14     &      7     \\
          &  93.90 \%  &   4.07 \%  &   2.03 \%  \\
\hline
Even-even &    155     &      3     &      1     \\
          &  97.48 \%  &   1.89 \%  &   0.63 \%  \\
\hline
Even-odd  &     69     &      2     &      1     \\
          &  95.83 \%  &   2.78 \%  &   1.39 \%  \\
\hline
Odd-even  &     60     &      4     &      2     \\
          &  90.91 \%  &   6.06 \%  &   3.03 \%  \\
\hline
Odd-odd   &     39     &      5     &      3     \\
          &  82.98 \%  &  10.64 \%  &   6.38 \%  \\
\hline\hline
Spherical &     68     &      0     &      0     \\
          & 100.00 \%  &   0.00 \%  &   0.00 \%  \\
\end{tabular}
\end{ruledtabular}
\end{table}

In Fig.~\ref{fig:compare}, the calculated results and experimental
values for $\alpha$ decay half lives are compared. In order to show
it clearly, these 344 nuclei are divided into four groups, namely,
159 even-even, 72 even-odd (even-$Z$ and odd-$N$), 66 odd-even, and
47 odd-odd nuclei. The ratios between the calculated and the
experimental values $S_\alpha = \log_{10}( T^\mathrm{Cal}_{1/2} /
T^\mathrm{Exp}_{1/2} )$ are presented in Fig.~\ref{fig:ratio}. Two
dashed lines are drawn to guide the eye. One finds that most of the
calculated results are of the same order of magnitude as the
experimental values.

A statistics of the agreement between the calculation and the
experiment is made and given in Table~\ref{tab:stati}. Among all
these 344 nuclei, there are only seven for which the calculated
$\alpha$ decay half lives deviate by more than two orders of
magnitude from the corresponding experimental values and 93.90\% of
them agree with experimental values within one order of magnitude.

Our results are particularly good for even-even nuclei, the
calculated half lives for 97.48\% of 159 even-even nuclei deviates
from the experiment by less than one order of magnitude. The ratio
$S_\alpha$ is less than one for 95.83 \% of 72 even $Z$ and odd $N$
nuclei, 90.91 \% of 66 odd-even nuclei and 82.98 \% of 47 odd-odd
nuclei. The angular momentum carried by the emitted $\alpha$
particle might not be zero for odd-$A$ or odd-odd nuclei. This will
bring in some errors for these nuclei in our calculation because the
centrifugal potential is ignored in the present study.

\begingroup
\begin{table}
\caption{\label{tab:spherical}Comparison between the calculated and
experimental $\alpha$ decay half lives of 68 nuclei of which the
daughter nuclei are spherical (with $|\beta_2|<0.01$). The charge
and mass numbers of the $\alpha$ decay nucleus are listed in the
first and the second columns. }
\begin{ruledtabular}
\begin{tabular}{ccc rr |ccc rr}
 $Z_\mathrm{p}$ & $A_\mathrm{p}$ & $Q_{ \alpha }$ & \multicolumn{2}{c}{$\log_{10}[T_{1/2}/\textrm{s}]$}
 &
 $Z_\mathrm{p}$ & $A_\mathrm{p}$ & $Q_{ \alpha }$ & \multicolumn{2}{c}{$\log_{10}[T_{1/2}/\textrm{s}]$}
 \\
    &     & (MeV)& Cal      & Exp      &    &     & (MeV) & Cal      & Exp      \\
 \hline
 52 & 106 & 4.30 & $ -3.83$ & $ -4.22$ & 84 & 207 &  5.22 & $  7.24$ & $  8.00$ \\
 60 & 144 & 1.91 & $ 23.17$ & $ 22.86$ & 84 & 208 &  5.22 & $  7.22$ & $  7.96$ \\
 61 & 145 & 2.32 & $ 17.53$ & $ 17.28$ & 84 & 210 &  5.41 & $  6.13$ & $  7.08$ \\
 62 & 146 & 2.53 & $ 15.61$ & $ 15.51$ & 84 & 212 &  8.95 & $ -6.58$ & $ -6.52$ \\
 62 & 148 & 1.99 & $ 23.63$ & $ 23.34$ & 84 & 213 &  8.54 & $ -5.58$ & $ -5.38$ \\
 63 & 147 & 2.99 & $ 11.32$ & $ 10.98$ & 84 & 214 &  7.83 & $ -3.62$ & $ -3.80$ \\
 64 & 148 & 3.27 & $  9.46$ & $  9.36$ & 84 & 215 &  7.53 & $ -2.71$ & $ -2.75$ \\
 64 & 150 & 2.81 & $ 13.91$ & $ 13.75$ & 84 & 216 &  6.91 & $ -0.61$ & $ -0.82$ \\
 66 & 150 & 4.35 & $  3.09$ & $  3.08$ & 84 & 218 &  6.11 & $  2.56$ & $  2.28$ \\
 66 & 152 & 3.73 & $  7.09$ & $  6.93$ & 85 & 213 &  9.25 & $ -6.95$ & $ -6.92$ \\
 68 & 152 & 4.94 & $  1.06$ & $  1.04$ & 86 & 200 &  7.05 & $  0.11$ & $  0.04$ \\
 68 & 154 & 4.28 & $  4.64$ & $  4.68$ & 86 & 202 &  6.78 & $  1.08$ & $  1.04$ \\
 70 & 154 & 5.47 & $ -0.33$ & $ -0.36$ & 86 & 203 &  6.63 & $  1.64$ & $  1.83$ \\
 72 & 156 & 6.04 & $ -1.66$ & $ -1.60$ & 86 & 204 &  6.55 & $  1.94$ & $  2.00$ \\
 72 & 158 & 5.41 & $  0.89$ & $  0.81$ & 86 & 206 &  6.39 & $  2.56$ & $  2.74$ \\
 74 & 158 & 6.60 & $ -2.76$ & $ -3.05$ & 86 & 207 &  6.25 & $  3.15$ & $  3.41$ \\
 82 & 210 & 3.79 & $ 16.16$ & $ 16.57$ & 86 & 208 &  6.26 & $  3.08$ & $  3.38$ \\
 84 & 190 & 7.64 & $ -2.51$ & $ -2.62$ & 86 & 214 &  9.21 & $ -6.52$ & $ -6.57$ \\
 84 & 191 & 7.48 & $ -2.03$ & $ -1.82$ & 86 & 215 &  8.84 & $ -5.63$ & $ -5.64$ \\
 84 & 192 & 7.32 & $ -1.53$ & $ -1.47$ & 86 & 216 &  8.20 & $ -3.93$ & $ -4.35$ \\
 84 & 193 & 7.10 & $ -0.80$ & $ -0.59$ & 86 & 217 &  7.89 & $ -3.04$ & $ -3.27$ \\
 84 & 194 & 7.00 & $ -0.47$ & $ -0.41$ & 86 & 218 &  7.26 & $ -1.02$ & $ -1.46$ \\
 84 & 195 & 6.75 & $  0.42$ & $  0.79$ & 87 & 215 &  9.54 & $ -6.94$ & $ -7.07$ \\
 84 & 196 & 6.66 & $  0.74$ & $  0.77$ & 87 & 217 &  8.47 & $ -4.32$ & $ -4.80$ \\
 84 & 197 & 6.41 & $  1.71$ & $  2.08$ & 88 & 216 &  9.53 & $ -6.59$ & $ -6.74$ \\
 84 & 198 & 6.31 & $  2.11$ & $  2.26$ & 88 & 218 &  8.55 & $ -4.17$ & $ -4.59$ \\
 84 & 199 & 6.08 & $  3.08$ & $  3.64$ & 88 & 220 &  7.60 & $ -1.35$ & $ -1.74$ \\
 84 & 200 & 5.99 & $  3.47$ & $  3.79$ & 89 & 217 &  9.83 & $ -6.93$ & $ -7.16$ \\
 84 & 201 & 5.81 & $  4.29$ & $  4.76$ & 89 & 219 &  8.83 & $ -4.56$ & $ -4.92$ \\
 84 & 202 & 5.70 & $  4.80$ & $  5.15$ & 90 & 218 &  9.85 & $ -6.65$ & $ -6.96$ \\
 84 & 203 & 5.50 & $  5.80$ & $  6.30$ & 90 & 220 &  8.95 & $ -4.51$ & $ -5.01$ \\
 84 & 204 & 5.49 & $  5.83$ & $  6.28$ & 91 & 219 & 10.09 & $ -6.85$ & $ -7.28$ \\
 84 & 205 & 5.32 & $  6.72$ & $  7.18$ & 91 & 221 &  9.25 & $ -4.93$ & $ -5.23$ \\
 84 & 206 & 5.33 & $  6.64$ & $  7.15$ & 92 & 222 &  9.50 & $ -5.20$ & $ -6.00$ \\
\end{tabular}
\end{ruledtabular}
\end{table}
\endgroup

The deformation influences the $\alpha$ decay life time both on the
preformation mechanism and on the penetration
process~\cite{Stewart1996a, Stewart1996, Denisov2005, Xu2006}. In
the present work, we have assumed the barrier potential to be
spherical. In 68 of these 344 nuclei, the spherical potential
assumption is met well (with $|\beta_2|<0.01$ for the daughter
nucleus~\cite{Raman2001}). In Table~\ref{tab:spherical} the
calculated and experimental values of the $\alpha$ decay half lives
for these nuclei are given. The statistical summary is also shown in
the last line of Table~\ref{tab:stati}. It is found that the new
formula gives very good results for these spherical nuclei. In most
cases, the differences between the calculated and the experimental
values of $\log_{10} T_{1/2}$ are smaller than 0.5. The root mean
square deviation between
$\log_{10}[T^\mathrm{Cal}_{1/2}/\textrm{s}]$ and
$\log_{10}[T^\mathrm{Exp}_{1/2}/\textrm{s}]$ is 0.34.

\section{Conclusion}
\label{sec:summary}

In the study of barrier penetration in nuclear physics, the
parabolic approximation is usually adopted because an analytical
solution exists for the penetrability of a parabola barrier
potential. The parabola approximation works indeed well both for the
penetrability and for the fusion cross section at energies around or
above the Coulomb barrier. But it fails at energies much smaller
than the barrier height due to the long-range Coulomb interaction.

In the present work, we derived a new barrier penetration formula,
Eq.~(\ref{eq:new}), based on the WKB approximation. We took into
account the influence of the long Coulomb tail in the barrier
potential properly. Therefore this formula is especially applicable
to the barrier penetration with penetration energy much lower than
the Coulomb barrier. We have shown that the present analytical
formula reproduces the WKB results very well.

This new penetration formula is used to calculate $\alpha$ decay
half-lives of 344 nuclei with the $\alpha$-nucleus potential given
in Ref.~\cite{Denisov2005}. Satisfactory agreement between the
present calculation and the experiment is achieved. For spherical
and even-even nuclei, the results are particularly good. Therefore,
the new formula could be used in the study of barrier penetration at
energies much smaller than the barrier height. Furthermore, we
expect that the new formula will facilitate the study of the barrier
penetrability where one has to introduce an energy-dependent
one-dimensional potential barrier or a barrier distribution
function.

\begin{acknowledgments}
This work was partly supported by the National Natural Science
Foundation (10575036, 10705014, and 10875157), the Major State Basic
Research Development Program of China (2007CB815000), the Knowledge
Innovation Project of CAS (KJCX3-SYW-N02 and KJCX2-SW-N17), and
Deutsche Forschungsgemeinschaft. The computation of this work was
supported by Supercomputing Center, CNIC, CAS.
\end{acknowledgments}

\begin{appendix}

\section{Derivation of the new penetration formula}
In order to evaluate the integration $x_2$ in Eq.~(\ref{eq:x1x2}),
we divide the potential between the position of the barrier
$R_\mathrm{B}$ and the outer turning point $R_\mathrm{out}$ into two
parts, $R_\mathrm{B} \le R \le R_\mathrm{V}$ and $R_\mathrm{V} \le R
\le R_\mathrm{out}$. $R_\mathrm{V}$ should be large enough so that
the nuclear potential vanishes for $R \ge R_\mathrm{V}$. For S wave,
\begin{eqnarray}
 x_2
 & = &
 2 \int^{R_ \mathrm{out} }_{R_\mathrm{B}}
    \sqrt{ \frac{2\mu }{ \hbar^2 } \left( V_\mathrm{N}(R) + V_\mathrm{C}(R) - E \right) }\ dR
 \nonumber \\
 & = &
 2 \int^{R_ \mathrm{V} }_{R_\mathrm{B}}
    \sqrt{ \frac{2\mu }{ \hbar^2 } \left( V_\mathrm{N}(R) + V_\mathrm{C}(R) - E \right) }\ dR
 \nonumber \\
 &   & \mbox{} +
 2 \int^{R_ \mathrm{out} }_{R_\mathrm{V}}
    \sqrt{ \frac{2\mu }{ \hbar^2 } \left( V_\mathrm{C}(R) - E \right) }\ dR
 \nonumber \\
 & = &
 2 \int^{R_ \mathrm{V} }_{R_\mathrm{B}}
    \frac{ \sqrt{ 2\mu } } { \hbar }
    \sqrt{ V_\mathrm{C}(R) - E }
    \sqrt{ 1 + \frac{ V_\mathrm{N}(R) }{ V_\mathrm{C}(R) - E } }\ dR
 \nonumber \\
 &   & \mbox{} +
 2 \int^{R_ \mathrm{out} }_{R_\mathrm{V}}
    \sqrt{ \frac{2\mu }{ \hbar^2 } \left( V_\mathrm{C}(R) - E \right) }\ dR
 .
\end{eqnarray}
It has been verified that when $R_\mathrm{V}$ is not very close to
$R_\mathrm{out}$, $| V_\mathrm{N}(R) / (V_\mathrm{C}(R) - E) | \ll
1$, therefore,
\begin{eqnarray}
 x_2
 & \approx &
 2 \int^{R_ \mathrm{V} }_{R_\mathrm{B}}
    \frac{ \sqrt{ 2\mu } } { \hbar }
    \sqrt{ V_\mathrm{C}(R) - E }
    \left[ 1 + \frac{1}{2} \frac{ V_\mathrm{N}(R) }{ V_\mathrm{C}(R) - E } \right]\ dR
 \nonumber \\
 &   & \mbox{} +
 2 \int^{R_ \mathrm{out} }_{R_\mathrm{V}}
    \sqrt{ \frac{2\mu }{ \hbar^2 } \left( V_\mathrm{C}(R) - E \right) }\ dR
 \nonumber
 \end{eqnarray}
 \begin{eqnarray}
 & = &
 2 \int^{R_ \mathrm{out} }_{R_\mathrm{B}}
    \frac{ \sqrt{ 2\mu } } { \hbar }
    \sqrt{ V_\mathrm{C}(R) - E }  \ dR
 \nonumber \\
 &   &
 +
   \int^{R_ \mathrm{V} }_{R_\mathrm{B}}
    \frac{ \sqrt{ 2\mu } } { \hbar }
    \frac{ V_\mathrm{N}(R) }{\sqrt{ V_\mathrm{C}(R) - E } }\ dR
 .
 \label{eq:x1x2_approx}
\end{eqnarray}
Since the Coulomb potential outside the barrier ($R\ge
R_\mathrm{B}$) is well described by [c.f. Eq.~(\ref{eq:Coulomb}],
\begin{equation}
 V_\mathrm{C}(R) = \frac{Z_1Z_2 e^2}{R},
 \label{eq:Coul}
\end{equation}
the first term in the above equation can be evaluated easily as,
\begin{eqnarray}
 x^{(1)}_2
 & \equiv &
 2 \int^{R_ \mathrm{out} }_{R_\mathrm{B}}
    \frac{ \sqrt{ 2\mu } } { \hbar }
    \sqrt{ V_\mathrm{C}(R) - E }  \ dR
 \nonumber \\
 & = &
 2 k R_\mathrm{B}
 \left[
  \tau \left( \frac{\pi}{2} - \arcsin
  \sqrt{ \frac{1}{\tau} }
  \right) -
  \sqrt{\tau - 1}
 \right]
 ,
\end{eqnarray}
with $k=\sqrt{2\mu E} / \hbar$ and $\tau =
V_\mathrm{C}(R_\mathrm{B}) / E$. For the evaluation of the second
term in Eq.~(\ref{eq:x1x2_approx}), we adopt a Woods-Saxon form for
the nuclear part of the barrier potential,
\begin{eqnarray}
 V_\mathrm{N}(R)
 & = & \frac{V_0}{1+\exp\left[(R-R_0)/a\right]}
 ,
\end{eqnarray}
and replace $\sqrt{ V_\mathrm{C}(R) - E }$ in the denominator by
$\sqrt{ V_\mathrm{C}(R_\mathrm{B}) - E }$,
\begin{eqnarray}
 x^{(2)}_2
 & \equiv &
   \int^{R_ \mathrm{V} }_{R_\mathrm{B}}
    \frac{ \sqrt{ 2\mu } } { \hbar }
    \frac{ V_\mathrm{N}(R) }{\sqrt{ V_\mathrm{C}(R) - E } }\ dR
 \nonumber \\
 & \approx &
   \int^{R_ \mathrm{V} }_{R_\mathrm{B}}
    \frac{ \sqrt{ 2\mu } } { \hbar }
    \frac{ V_\mathrm{N}(R) }{\sqrt{ V_\mathrm{C}(R_\mathrm{B}) - E } }\ dR
 \nonumber \\
 & = &
 \frac{ k } { \sqrt{ \tau - 1 } } \frac{V_0}{E}
 \left. \left\{
 R - a \ln [ 1 + e^{ (R - R_0) / a } ]
 \right\} \right|_{R_B}^{ R_{ \mathrm{V} } }
 \nonumber \\
 & \approx &
 \frac{ k } { \sqrt{ \tau - 1 } } \frac{V_0}{E}
 \{
 R_0- R_B + a \ln [ 1 + e^{ (R_B - R_0) / a } ]
 \}
 \nonumber \\
 & = &
 \frac{ k a } { \sqrt{ \tau - 1 } } \frac{V_0}{E}
 \ln [ 1 + e^{ (R_0 - R_B) / a } ]
 .
\end{eqnarray}
In the above derivation, we have used the fact that
$\exp\left[(R_\mathrm{V}-R_0)/a\right] \gg 1$ for $\alpha$ decay and
penetration well below the Coulomb barrier. Finally, we have an
analytical expression for $x_2$,
\begin{eqnarray}
 x_2
 & = &
 2 k R_\mathrm{B}
 \left[
  \tau \left( \frac{\pi}{2} - \arcsin
  \sqrt{ \frac{1}{\tau} }
   \right) -
  \sqrt{\tau - 1}
 \right]
 \nonumber \\
 &   & \mbox{}
 +
 \frac{ k a } { \sqrt{ \tau - 1 } } \frac{V_0}{E}
 \ln [ 1 + e^{ (R_0 - R_B) / a } ]
 .
\end{eqnarray}

\end{appendix}

\bibliographystyle{apsrev}

\end{document}